\documentclass[useAMS,usenatbib]{mn2e}
\usepackage{amsmath}
\usepackage{graphicx}
\usepackage{epstopdf}
\usepackage{amssymb}
\usepackage{extarrows}
\usepackage{color}
\usepackage{amsmath,bm}
\usepackage{CJK}

\newcommand{\be}{\begin{equation}}
\newcommand{\ee}{\end{equation}}
\newcommand{\gr}{\mathrm{GR}}

\newcommand{\oct}{\mathrm{oct}}

\newcommand{\au}{\mathrm{au}}

\newcommand{\hill}{\mathrm{Hill}}
\newcommand{\msun}{\mathrm{M_{\sun}}}
\newcommand{\pc}{\mathrm{pc}}
\newcommand{\yr}{\mathrm{yr}}
\newcommand{\ev}{\mathrm{EV}}
\newcommand{\rel}{\mathrm{REL}}
\newcommand{\rrs}{\mathrm{RRS}}
\newcommand{\rrv}{\mathrm{RRV}}
\newcommand{\kms}{\mathrm{kms^{-1}}}
\newcommand{\kb}{\mathrm{K,B}}
\newcommand{\kc}{\mathrm{K,C}}
\newcommand{\np}{\mathrm{NP}}
\def\e1{e_1^2}
\def\I{i_{\mathrm{tot}}}

\title[Stellar binaries surrounding SMBH binary]{Modified evolution of stellar binaries from supermassive black hole binaries}

\author[Liu, Wang \& Yuan]{Bin Liu$^{1,2,3}$,
Yi-Han Wang$^{2,3}$  and Ye-Fei Yuan$^{2,3}$\\
$^{1}$ Shanghai Astronomical Observatory, Chinese Academy of Sciences, 80 Nandan Road, Shanghai 200030, China\\
$^{2}$ Department of Astronomy, University of Science and Technology of China, Hefei, Anhui 230026, China\\
$^{3}$ CAS Key Laboratory for Research in Galaxies and Cosmology, Hefei, Anhui 230026, China
}

\begin{document}


\pagerange{\pageref{firstpage}--\pageref{lastpage}} \pubyear{2014}

\maketitle

\label{firstpage}

\begin{abstract}
{
The evolution of main sequence binaries resided
in the galactic centre is influenced a lot by the central super massive black hole (SMBH).
Due to this perturbation, the stars in a dense environment are likely to experience mergers or collisions through
secular or non-secular interactions.
In this work, we study the dynamics of the stellar binaries at galactic center, perturbed by another distant SMBH.
Geometrically, such a four-body system is supposed to be decomposed
into the inner triple (SMBH-star-star) and the outer triple (SMBH-stellar binary-SMBH).
We survey the parameter space and determine the criteria analytically for the stellar mergers and the tidal disruption events (TDEs).
For a relative distant and equal masses SMBH binary, the stars have more opportunities to merge
as a result from the Lidov-Kozai(LK) oscillations in the inner triple.
With a sample of tight stellar binaries,
our numerical experiments reveal that a significant fraction of the binaries,
$\sim70\%$, experience merger eventually.
Whereas the majority of the stellar TDEs are likely to
occur at a close periapses to the SMBH, induced by the outer Kozai effect.
The tidal disruptions are found numerically as many as $\sim10$ per cent 
for a close SMBH binary
that is enhanced significantly than the one without the external SMBH.
These effects require the outer perturber to have an inclined orbit ($\geq40^\circ$) relatively to the inner orbital plane
and may lead to a burst of the extremely astronomical events associated with the detection of the SMBH binary.
}
\end{abstract}


\section{Introduction}

The supermassive black hole (SMBH) resided in the galactic centre is ubiquitous \citep[e.g.,][]{Ho}.
At present, in the inner zone of our Milky Way, within $\sim$ 1 pc, the
stars are observed in a dense stellar environments, where the gravitational potential
is dominated by the central SMBH \citep[e.g.,][]{Alexander}.
Since the high density of stars in the galactic centre,
the majority of stars in the field, are believed to reside in binaries or higher multiplicity systems \citep[e.g.,][]{Raghavan,Evans}.
Furthermore, these binaries are gravitational bound to the SMBH
and effectively form hierarchical triple systems with the SMBH.

Therefore, it is interesting to take account a close encounter between a stellar binary and an SMBH.
When the centre of mass of the binary approaches the SMBH with the pericentre distance below
the tidal break-up of the binary, both the angular momentum and the orbital energy of the binary are exchanged
with those of the SMBH.
Recently, the evolution of the binary orbiting around the SMBH with the time-scale shorter than the secular times-cale has been studied extensively:
(i) when the stellar binaries undergo low angular momentum orbits, the stars are considered to be unbound to each other by the strong tidal field
of the SMBH. As a result, one component is captured by the SMBH, while
the other is ejected with high velocity on the order of $500-1000$ km s$^{-1}$, which may
be responsible for the hypervelocity stars (HVSs)
\citep[e.g.,][]{Hills H,Yuqingjuan,Antonini 09};
(ii) due to the high density of the stellar distribution at the galactic center,
stars in a binary approaching the SMBH can be tidally disrupted
if the relative distance is on the order of the tidal disrupted radius of a single star.
The accretion of the stellar debris results in a strong flare in the electromagnetic radiation,
leading to a tidal disruption event (TDE)
\citep[e.g.,][]{Hills T,Rees,Phinney,Gezari,DoubleTDE};
(iii) similarly, the interaction between the SMBH and the stellar binary may also lead to
a physical collision between two components of the binary
\citep[e.g.,][]{Ginsburg}.

If the orbital plane of the stellar binary is inclined with the external plane of the SMBH
over $40^\circ$, the stellar eccentricity and inclination will experience
periodic oscillations on the secular time-scale,
known as Lidov-Kozai (LK) oscillations \citep[e.g.,][]{Lidov,Kozai}.
In addition to the leading order (quadrupole) approximation,
it has been recognized that high-order (octupole) interactions
between the inner and outer orbits can lead to more abundant dynamical
behaviours in hierarchical triples \citep[e.g.,][]{Harrington,Marchal,Ford,Smadar 2011,SRF}.
The effects of LK oscillations in the formation and evolution of various
astrophysical systems have been extensively studied in recent years
\citep[e.g.,][]{Eggleton,FT,Holman,Innanen,WM,DongLai S,Resonance,Anderson}.
In particular, in the environment of SMBH, the formation and merger of
black hole binaries at the centres globular clusters or
galaxies is increased significantly
\citep[e.g.,][]{Blaes,MH,Wen,AMM},
as well as the production of
Type Ia supernova from white-dwarf binary mergers
\citep[e.g.,][]{Thompson,PMT}
or direct collisions \citep[e.g.,][]{Katz,Kushnir}.
Also,
an enhanced rate of stellar TDEs is studied in the SMBH binary,
due to the eccentric LK mechanism \citep[e.g.,][]{Li TDE}.

The motion of the binaries in the galactic centre is varied by the influence of the SMBH.
This process is supposed to be accelerated by another massive perturber outside this triple system.
In this paper, we consider the stellar binaries orbiting around an SMBH,
perturbed by a distant SMBH. As expected, the evolution of the stellar binary is modified by the
existence of the forth body, re-populating the stars around the first SMBH.
Different from the simple three-body system,
the existence of the second SMBH allows us to construct two triple systems.
If the ratio of distances between the stars and the SMBH is sufficient small,
and the mass of star is negligible compared to the SMBH,
the stellar binary can be treated as a test mass particle.
So, in addition to the stellar binary-SMBH system, the outer triple system
(SMBH-stellar binary-SMBH) is formed.
Obviously, when the inner triple has the mutual inclination lied in the range $40^\circ\sim140^\circ$,
the LK oscillations occur in this three body system.
The eccentricity of the stellar binary can be excited, leading to a merger between two components.
Inversely, when the Kozai effect of the outer triple come into play,
the stellar binary will have a eccentric orbit moving around the SMBH.
If the periastron is small sufficiently, the binary may be tidal break-up after the close passage,
producing TDE or HVS.
Therefore,
the dynamics in the latter case is expected to be rich and interesting, and we will focus on this regime in this work.

Our paper is organized as follows.
In Section 2, we introduce the geometry of the four-body system.
Through the comparison of different time-scales, we characterize the parameter space
where the outer LK oscillation is dominated.
In Section 3, we carry out the $N$-body simulations to explore the evolution of the stellar binary
approaching the SMBH.
Furthermore, considering an illustrative example, we study the dynamics of the four-body system
and obtain the rates of different events in Section 4.
We also trace the variations of the rates when the SMBH binary has a different semimajor axis.
Our results and conclusions are summarized in Section 5.

\section{Stellar Binary surround SMBH Binary}

\begin{figure}
\centering
\begin{tabular}{c}
\includegraphics[width=7cm]{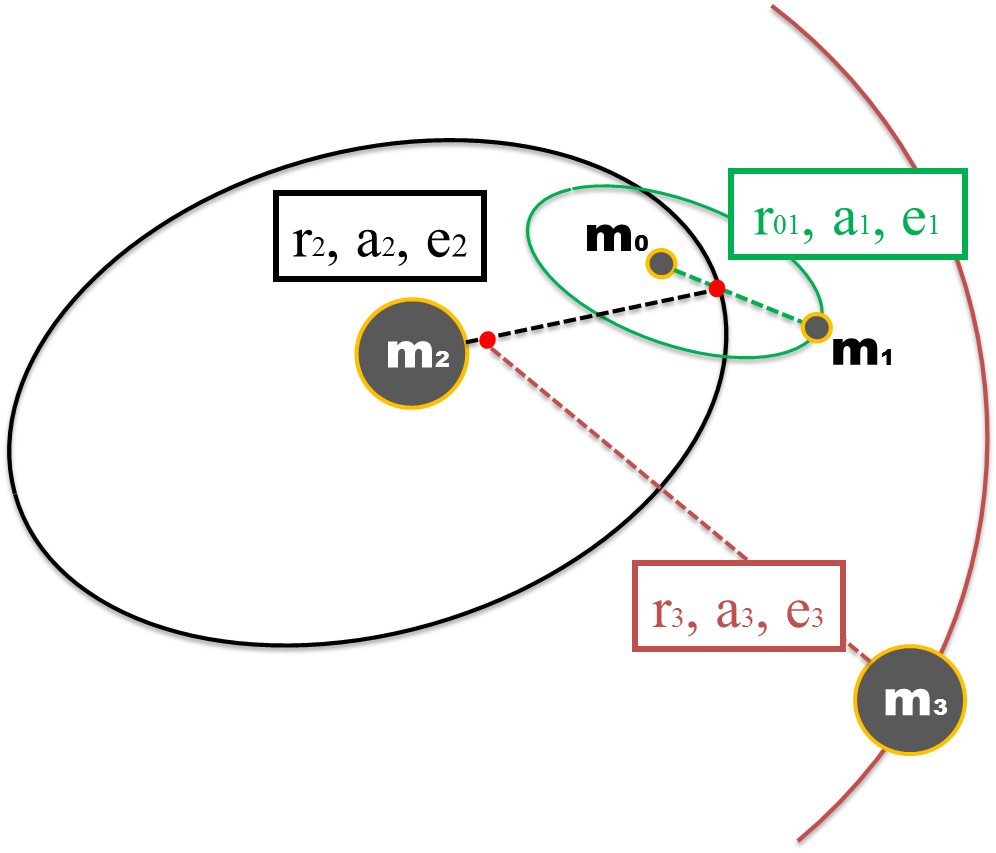}\\
\\
\\
\includegraphics[width=7cm]{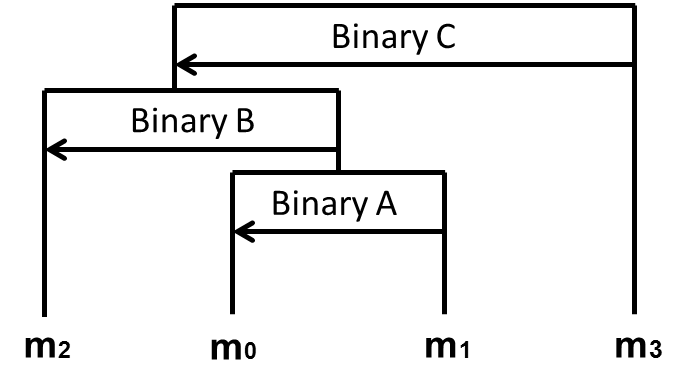}
\end{tabular}
\caption{The configuration of the four-body system we studied.
Binary A comprises two stars ($m_0$ and $m_1$);
Binary B consists of the Binary A and an SMBH ($m_2$),
while the outer SMBH ($m_3$) orbits the centre of mass of Binary B, constituting Binary C.
Here, $a_{1,2,3}$ is the semimajor axes, $e_{1,2,3}$ is the eccentricities
and $r_{01,2,3}$ is the separation between two components of each binary, respectively.
}\label{fig:1}
\end{figure}

We study the merger and the tidal break-up of the stellar binaries surrounding the
SMBH binary.
First, as shown in the upper panel of Fig. \ref{fig:1},
we introduce the masses of the stars by $m_0$ and $m_1$ (stellar binary components),
$m_2$ (the first SMBH), and $m_3$ (the second SMBH).
Then, for the orbital parameters we use $a_{1,2,3}$ as the semimajor axes, $e_{1,2,3}$ as the eccentricities
and $r_{01,2,3}$ as the separations between two components of each binary.
Subscript ``$1,2,3$" denote three binaries (Binary A, B and C, respectively), as shown in the lower panel of Fig. \ref{fig:1}.
Based on the configuration, this four-body system can be decomposed into two three-body systems:
the ``inner triple" is the stellar binary-SMBH triple system, consisting of $m_0-m_1-m_2$;
while the ``outer triple" is consisted of the stellar binary - SMBH binary system, i.e., $m_2-(m_0+m_1)-m_3$.

\subsection{Time Scales}

In order to identify the parameter space where Kozai effect is important,
we consider the different sources of apsidal precession in the galactic nuclei.
For the inner triple system, the stellar binaries are immersed in a dense environment.
Other gravitational process may come into play and affect the evolution of the stellar binary, even change the distribution.

Binaries orbited by a highly inclined perturber will undergo LK oscillations.
In the inner triple, the corresponding time-scale at the approximation of quadrupole level is described as
\citep[e.g.,][]{Kozai,Lidov}
\be\label{eq:KozaiB}
\begin{split}
T_\kb=&\frac{1}{n_1}\bigg(\frac{m_0+m_1}{m_2}\bigg)\bigg(\frac{a_2}{a_1}\bigg)^3(1-e_2^2)^{3/2}\\
=&1.3\times10^6 \yr\bigg(\frac{m_2}{1\times10^6\msun}\bigg)^{-1}\bigg(\frac{m_0+m_1}{1\msun}\bigg)^{1/2}\\
&\times\bigg(\frac{a_2}{0.1\pc}\bigg)^3\bigg(\frac{a_1}{1au}\bigg)^{-3/2}(1-e_2^2)^{3/2},
\end{split}
\ee
where $n_1=\sqrt{G(m_0+m_1)/a_1^3}$ is the mean motion of the stellar binary.

Compared with the Kozai time-scale, some related time-scales are reviewed as follows.
First, to date, we have only little understanding about the distribution of the
main sequence stars resided in the galactic centre from observation \citep[e.g.,][]{Duquennoy and Mayor,Willems and Kolb},
whereas most of the theoretical studies are focused on the close binary (i.e., X-ray binary and gravitational wave source)
and the distribution predicted has a strong dependence on the models.
For simplicity, in the following text,
we take an assumption that the stellar density profile is described as a power law
and normalized at $1$ pc \citep[e.g.,][]{Alexander,Schodel,Antonini 12} as
\be\label{eq:density}
\rho(r)=\rho_0\bigg(\frac{r}{r_0}\bigg)^{-\alpha}\bigg[1+\bigg(\frac{r}{r_0}\bigg)^2\bigg]^{(\alpha-1.8)/2},
\ee
where $\rho_0=5.2\times10^5\msun \pc^{-3}$, $r_0=0.5\pc$.
The configuration of stars observed at the galactic nuclei is given when the inner slope index $\alpha=0.5$.
Here, we assume $\alpha$ to be $1.8$, representative of the relaxed distribution of stars dominated by a point-mass potential.

Near the galactic center, the kinetic energy of the field stars may be larger than that of the stellar binary,
where the binary might evaporate due to the sufficient encounters with the surrounding stars.
The associated time-scale is given by \citep[e.g.,][]{Binney and Tremaine}
\be
\begin{split}\label{eq:TEV}
T_{\ev}=&1.6\times10^7\yr \bigg(\frac{m_0+m_1}{1\msun}\bigg)\bigg(\frac{\sigma}{306~\kms}\bigg)\\
&\times\bigg(\frac{a_1}{1\au}\bigg)^{-1}\bigg(\frac{\ln \Lambda}{15}\bigg)^{-1}\bigg(\frac{\rho}{2.1\times10^6\msun}\bigg)^{-1},
\end{split}
\ee
where $\ln \Lambda\sim\ln[m_2/(m_0+m_1)]$ is the Coulomb logarithm.
Here, the one-dimensional velocity dispersion, $\sigma$, is satisfied with the Jeans equation
\be
\rho(r)\sigma(r)^2=G\int^\infty_r\frac{\rho(r')}{r'^2}\big[m_2+M_\star(r')\big]dr',
\ee
with $M_\star(r')$ the mass of the stars within the radius $r$.

The encounters between the stellar stars may also push the galaxy to evolve into equilibrium,
leading to dynamical relaxation. Supposing the equal mass stars have distribution with isotropic velocity,
the time-scale is defined as \citep[e.g.,][]{Spitzer}
\be
\begin{split}
T_\rel=&1.6\times10^{10}\yr \bigg(\frac{\sigma}{306~\kms}\bigg)
\bigg(\frac{\ln \Lambda}{15}\bigg)^{-1}\\
&\times\bigg(\frac{\rho}{2.1\times10^6\msun}\bigg)^{-1}.
\end{split}
\ee

In the dense environment, such as the stellar cluster,
the angular momentum of the orbit of the stellar binary around SMBH may change 
both the magnitude and the orientation by
the resonant relaxation. Here, we take into account the time-scale due to the scalar resonant relaxation (related to magnitude) as
\citep[e.g.,][]{Rauch and Tremaine}
\be
T_\rrs=4.6\times10^{8}\yr \bigg(\frac{m_2}{1\times10^6\msun}\bigg)^{1/2}
\bigg(\frac{a_2}{0.1\pc}\bigg)^{3/2},
\ee
and the vector resonant relaxation (relevant for orientation) as \citep[e.g.,][]{Merritt}
\be\label{eq:RRV}
\begin{split}
T_\rrv=&3.8\times10^{6}\yr \bigg(\frac{m_2}{1\times10^6\msun}\bigg)^{1/2}
\bigg(\frac{a_2}{0.1\pc}\bigg)^{3/2}\\
&\times\bigg(\frac{N}{6000}\bigg)^{-1/2},
\end{split}
\ee
where $N$ is the number of the stars interior to the semimajor axis $a_2$.

If the stellar binary separation at pericentre is sufficiently small,
the additional forces, such as general relativity (GR) may overcome the tidal torque exerted by the
outer binary, suppressing the excitation of the eccentricity of the stellar binary
\citep[e.g.,][]{Blaes,Smadar 2013a}.
The time-scale of the precession of the argument of periapsis caused by the first order post Newtonian (PN) correction
in the inner and outer orbit is
\be\label{eq:TGRB1}
\begin{split}
T_{\gr,\mathrm{B1}}&=\frac{2\pi c^2}{3G^{3/2}}\frac{a_1^{5/2}(1-e_1^2)}{(m_0+m_1)^{3/2}}\\
&=3.4\times10^7 \yr \bigg(\frac{a_1}{1\au}\bigg)^{5/2} \bigg(\frac{m_0+m_1}{1 \msun}\bigg)^{-3/2}\\
&~~~~\times(1-e_1^2),
\end{split}
\ee
and
\be\label{eq:TGRB2}
\begin{split}
T_{\gr,\mathrm{B2}}&=\frac{2\pi c^2}{3G^{3/2}}\frac{a_2^{5/2}(1-e_2^2)}{(m_0+m_1+m_2)^{3/2}}\\
&=2.1\times10^9 \yr \bigg(\frac{a_2}{0.1\pc}\bigg)^{5/2} \bigg(\frac{m_0+m_1+m_2}{1\times10^6 \msun}\bigg)^{-3/2}\\
&~~~~\times(1-e_2^2),
\end{split}
\ee
respectively. In particular, the interaction between the inner and outer binaries at 1 PN order
can affect the long-term evolution of the triple. The related time-scale is given by
\citep[e.g.,][]{Smadar 2013a,Clifford 1,Clifford 2}
\be\label{eq:TGRB3}
\begin{split}
T_{\gr,\mathrm{B3}}&=\frac{16}{9}\frac{c^2}{G^{3/2}}\frac{a_2^3}{a^{1/2}}\frac{(1-e_2^2)^{3/2}}{e_1(1-e_1^2)^{1/2}}
\frac{(m_0+m_1)^{3/2}}{(m_0^2+m_0 m_1+m_1^2)m_2}\\
&=2.5\times10^{14} \yr \bigg(\frac{a_2}{0.1\pc}\bigg)^{3} \bigg(\frac{a_1}{1\au}\bigg)^{-1/2}
\bigg(\frac{m_0+m_1}{1 \msun}\bigg)^{3/2}\\
&~~~~\times\bigg(\frac{m_0^2+m_0 m_1+m_1^2}{1 \msun}\bigg)^{-1}
\bigg(\frac{m_2}{1\times10^6 \msun}\bigg)^{-1}\\
&~~~~\times\frac{(1-e_2^2)^{3/2}}{e_1(1-e_1^2)^{1/2}}.
\end{split}
\ee

\begin{figure}
\centering
\includegraphics[width=8.5cm]{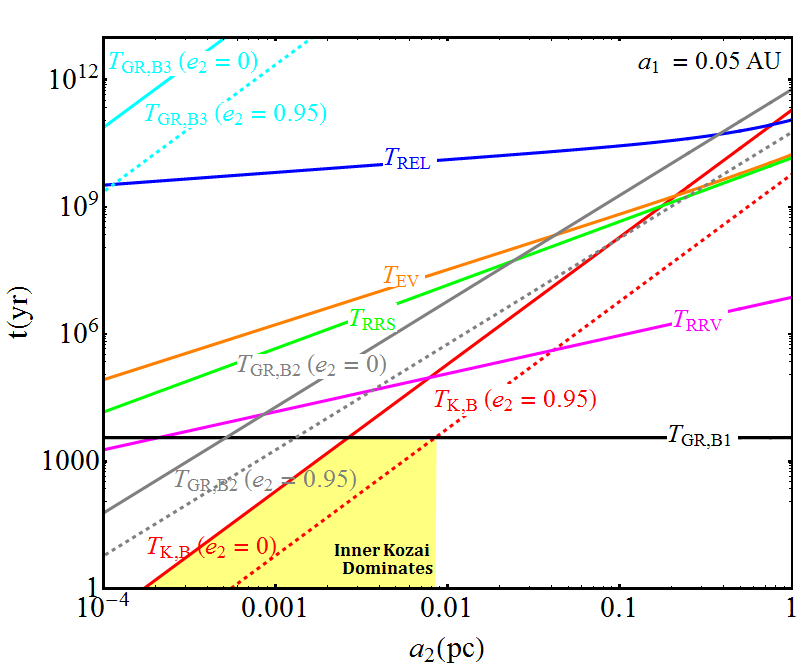}
\includegraphics[width=8.5cm]{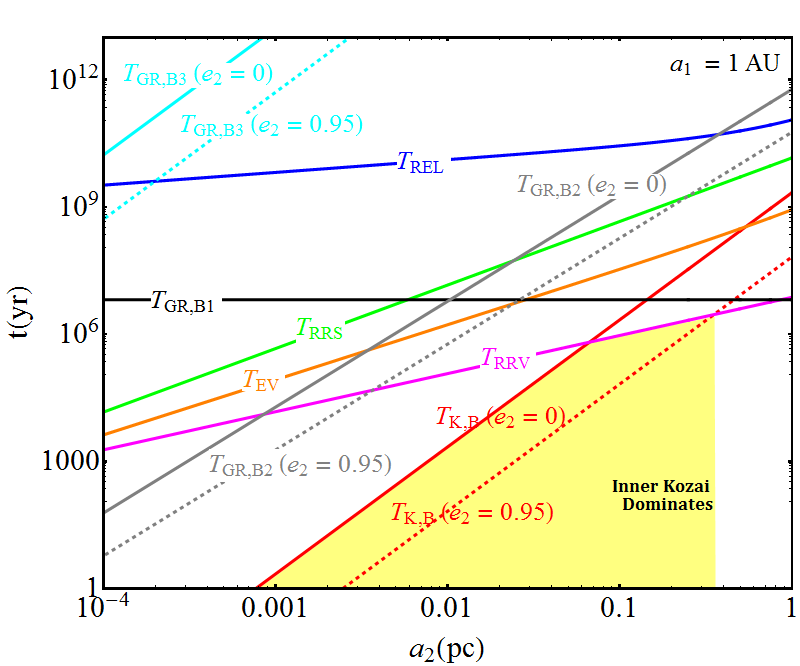}
\caption{Different time-scales associated with the galactic centre and the LK oscillations
(equations \ref{eq:TEV}-\ref{eq:TGRB3})
as a function of $a_2$, for an illustrative example of $m_0=2 \msun$, $m_1=1 \msun$, $m_2=10^6 \msun$ and $e_{1,0}=0.001$.
In the upper panel, $a_1$ is assumed to be $0.05\au$,
and in the lower panel $a_1$ equals to $1\au$.
}\label{fig:2}
\end{figure}

Figure \ref{fig:2} shows a comparison of the time-scales for the illustrative example:
$m_0=2 \msun$, $m_1=1 \msun$, $m_2=10^6 \msun$ and $e_{1,0}=0.001$.
In the upper panel, the stellar binary has the small semimajor axis as $a_1=0.05\au$
(requiring to be survival after many orbits around the SMBH in the further simulations),
while $a_1=1\au$ in the lower panel.
At $\gtrsim0.01\pc$, Kozai effect is completely quenched by the GR effect for the much tighter orbit
$a_1=0.05\au$ with fixed $e_2$.
But the boundary of Kozai-dominated region of the inner triple is shifted to $\sim0.4\pc$ for the larger $a_1$. Typically,
compared with other time-scales, there is an upper limit on $a_2$ for a fixed $a_1$.
In fact, the observed stellar disc also has the radial extended from $\sim0.04$pc
to $\sim4$pc \citep[]{Alexander, Prodan}.
Thus, we identify the parameter space with moderately small $a_2$ in the following
test, irrespective of other dynamics.

\subsection{Parameter space of outer Kozai Oscillations}

\begin{figure*}
\centering
\includegraphics[width=15cm]{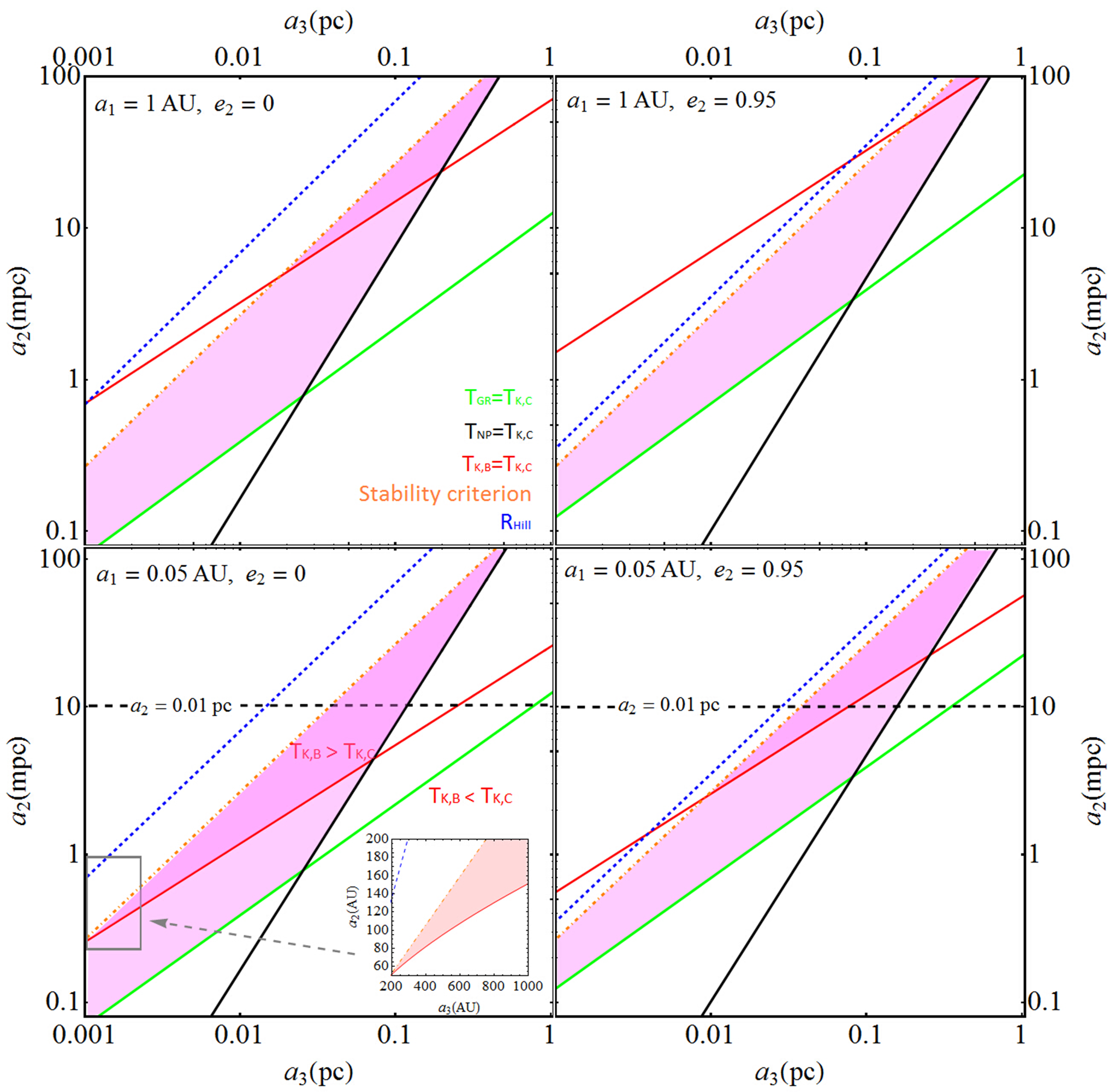}
\caption{The $a_2-a_3$ parameter space where the outer LK oscillation is dominated.
Here, we choose the parameters as $m_0=2 \msun$, $m_1=1 \msun$, $m_2=m_3=10^6 \msun$ and $e_{1,0}=e_{3,0}=0.001$.
Different values of $a_1$ and $e_2$ are included as labeled.
The parameter space bounded by the stability criterion, $T_\gr=T_\kc$ and $T_\np=T_\kc$,
corresponds to the Kozai effect allowed in the outer triple (red region).
But it may dominate the dynamics of the system with the parameters above the curve of $T_\kb=T_\kc$.
The dashed line of $a_2=0.01\pc$ marked in the lower panels give the upper limit on $a_2$,
above which the inner LK oscillations are quenched by other processes related in the galactic nuclei (see Fig. \ref{fig:2}).
}\label{fig:3}
\end{figure*}

As shown in Fig. \ref{fig:2},
the increased eccentricity of Binary B would decrease the Kozai time-scale at the quadrupole level,
and does not affect others.
That is to say,
the highly eccentric orbit has more opportunities to affect the evolution of the stellar binary near the galactic nuclei,
which can be produced by the scattering between stars \citep[e.g.,][]{Rauch and Tremaine,Perets}
or by the Kozai oscillations induced by a massive and distant perturber, such as another SMBH.

The LK mechanism in the outer triple operates if the following criteria are satisfied:

(i) The stellar binary (Binary A) stays in the Hill sphere of the inner SMBH ($m_2$)
in order for them to remain bound to it
\be\label{eq:Hill}
R_\hill=a_3(1-e_3)\bigg(\frac{m_2}{3m_3}\bigg)^{1/3}>a_2(1+e_2).
\ee

(ii) The configuration of Binary C should be satisfied with the hierarchical condition
\be\label{eq:varepsilon}
\varepsilon_\oct=\frac{a_2}{a_3}\frac{e_3}{1-e_3^2}<0.1,
\ee
which is always valid when $e_3=0$.

(iii) The conditions of parameters of Binary C are subject to the stability criterion of \citet{MA}
\be\label{eq:MA}
\frac{a_3}{a_2}>2.8\bigg(1+\frac{m_3}{m_0+m_1+m_2}\bigg)^{2/5}
\frac{(1+e_3)^{2/5}}{(1-e_3)^{6/5}}.
\ee

(iv) The quadrupole Kozai time-scale of the outer triple
\be\label{eq:KozaiC}
T_\kc=\frac{1}{n_2}\bigg(\frac{m_0+m_1+m_2}{m_3}\bigg)\bigg(\frac{a_3}{a_2}\bigg)^3(1-e_3^2)^{3/2},
\ee
should be smaller than the time-scale of 1 PN corrections in Binary B
\be\label{eq:TGRC1}
T_{\gr,\mathrm{C1}}=T_{\gr,\mathrm{B2}}=\frac{2\pi c^2}{3G^{3/2}}\frac{a_2^{5/2}(1-e^2_2)}{(m_0+m_1+m_2)^{3/2}},
\ee
and the Newtonian precession time-scale associated with the precession of the stellar binary due to the gravitational potential
of stars around the first SMBH \citep[e.g.,][]{Kocsis, Li TDE}
\be\label{eq:TNP}
T_\np=2\pi\bigg(\frac{n_2}{\pi m_2 e_2}\bigg|\int_0^\pi M_\star(r)\cos\psi d\psi\bigg|\bigg)^{-1},
\ee
where $n_2=\sqrt{G(m_0+m_1+m_2)/a_2^3}$ is the mean motion of Binary B and
$r=r(\psi)=a_2(1-e_2^2)/(1+e_2\cos\psi)$ in Keplerian motion.

To make the Kozai effect prominent in the outer triple,
it is required to compare the double Kozai time-scales of the inner and outer triples.
If the oscillating period driven by the inner triple is much shorter and the one from the outer,
the Kozai mechanism may play an important role in the inner triple (Binary A and an SMBH).
The eccentricity of Binary A can be driven to approach unit, leading to the merger event between $m_0$ and $m_1$.
Meanwhile, the perturbation from the second SMBH can be neglected ($e_2$ will not change so much).
Conversely, the evolution of the four-body system will be
dominated over by the LK oscillations in the outer triple (Binary B and the second SMBH)
if the mutual inclination lies in the range of $40^\circ-140^\circ$
and the above conditions are satisfied. It is noted that the inner LK oscillations are still allowed to occur
when $T_\kb>T_\kc$, but the dynamics of the system is dominated by the outer one.
Therefore, $e_2$ can be excited faster than $e_1$, making Binary A highly elliptical bound orbit
(even a parabola), whose pericentre lies close to the SMBH ($m_2$).
That means, the stellar binary might undergo tidal separation and TDEs or HVSs would occur in succession.

Considering the previous illustrative example, we include a second SMBH ($m_3=10^6\msun$) and
start the exploration in $a_2-a_3$ parameter space (as shown in Fig. \ref{fig:3}).
The initial eccentricities of the binaries are chosen to be zero ($e_{1,0}=e_{3,0}=0$),
and two semimajor axes of the stellar binary are considered to include $a_1=0.05\au$ and $a_1=1\au$ (upper and lower panels).
Here, the distant unit is pc, which makes a large room for the competition of the double Kozai oscillations.
In addition, we zoom in the small-scale area in the unit of AU.
To test the dependence on $e_2$,
we assume $e_2=0$ (left-hand panels) and $e_2=0.95$ (right-hand panels), respectively.
As seen, the outer LK oscillations are allowed to occur in the red region, bounded by the criteria
(equations \ref{eq:MA}-\ref{eq:TNP}).
However, the corresponding parameter space is divided into two parts by the curve of $T_\kb=T_\kc$,
and left side denotes $T_\kb>T_\kc$ while opposite in the right side.
The outer Kozai dominated region is reduced into a narrower space (deep red region).
On the other hand, requiring the inner Kozai time-scale to be shorter than GR effects,
the compact stellar binary ($a_1=0.05\au$)
cannot be so far away from SMBH, and the region of $T_\kb<T_\kc$ should be limited at
$a_2\sim0.01$ pc. It is also required to note that the parameter space nearby the curve of $T_\kb=T_\kc$
is going to be governed by the inner triple with high $e_2$.
In all, we find that the outer Kozai dominated region becomes relatively larger with a small $a_1$,
although the physical processes are the same.

\section{Basic Model of Tidal Break-Up of Binaries}

\begin{figure}
\centering
\includegraphics[width=7cm]{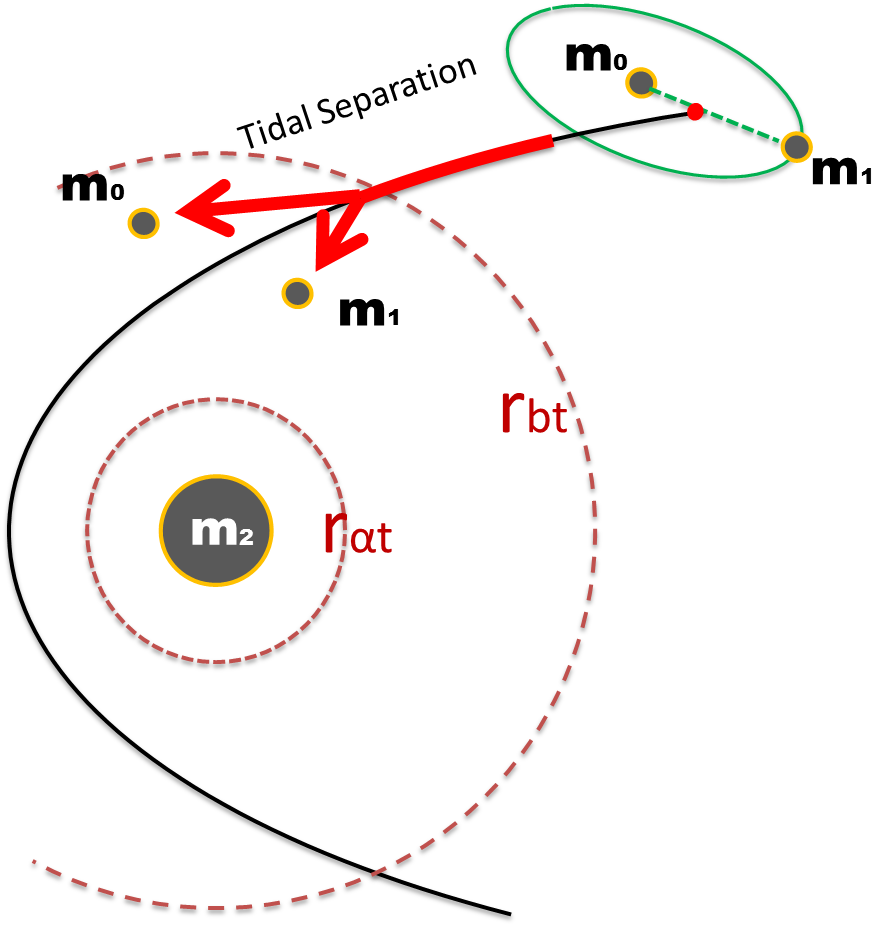}
\caption{
The illustration of the motion of the stellar binary with a highly elliptical orbit,
whose pericentre is close to the SMBH.
When the angular momentum of the stellar binary is sufficient small,
entering `Tidal Loss Cone', the centre of mass of the binary crosses the tidal separated radius ($r_{\rm bt}$) and the orbit
becomes unstable, resulting unbound stars.
Since the orbital energy is exchanged during the close passage,
the subsequent evolution may be one of the mergers, TDE (accompanied by HVS) or
surviving.
}\label{fig:4}
\end{figure}

Before running the simulations of the four-body system systematically,
we explore the close periapse approach of a binary in a highly eccentric orbit
around the SMBH.
As shown in Fig. \ref{fig:4},
when $r_2$, the separation between the centre of 
the mass of Binary A and the SMBH, is below the tidal separated radius of the binary:
\be\label{eq:RBT}
r_{bt}\sim a_1\Big(\frac{m_2}{m_0+m_1}\Big)^{1/3},
\ee
the binary is considered to be broken apart by the tidal field.
Since the orbital energy is exchanged,
the dynamics of the inner three-body may become chaotic during the close passage by the first SMBH.
Following the tidal separation of the binary, both stars remain close to the original Keplerian orbits of the
binary around the SMBH, separately.
Then, the stars will either scatter off or undergo individual TDE immediately.
Here, we introduce the tidal radius of each star as
\be\label{eq:RAT}
r_{\alpha t}= 3 R_\alpha\Big(\frac{m_2}{m_\alpha}\Big)^{1/3}~(\alpha=0,1),
\ee
where $R_\alpha$ is the stellar radius and the coefficient $3$ refer to the
case where the star will experience accumulated heating to be disrupted outside the ``pure" tidal radius \citep[]{Li Loeb}.
We assume a mass-radius relation \citep[e.g.,][]{Hansen} as
\be
R_\alpha=R_{\sun}\bigg(\frac{m_\alpha}{\msun}\bigg)^{3/4},
\ee
where $R_{\sun}$ is radius of the sun.

\subsection{Numerical setup}

In this work, all the dynamics are carried out by using a c++ code with explicit Frog Leaping Algorithm.
The method includes the adaptive step with error determined by the relative position that set to be $10^{-9}$ to trace the motion of binaries accurately in
arbitrarily mass ratios. To test our code, we reproduce Fig. 10 in \cite{Smadar 2013b} and find perfect agreement.
The code can be downloaded freely at https://github.com/wyh102324/SMBHbinary

In this section, for simplicity,
we place the stellar binary ($m_0=2 \msun$, $m_1=1 \msun$ and $a_1=0.05\au$) at the apastron of the
outer orbit ($m_2=10^6 \msun$) initially, outside the binary tidal separated radius.
The initial orbital phase is fixed, i.e, the arguments of periapse of the two orbits defer n$\pi$ (n is a integer).
But the eccentricity $e_2\in(0,1)$, semimajor axis $a_2\in(1,250)\au$ and the inclination between the orbital planes of
Binaries A and B $\I\in(0^\circ,180^\circ)$ are distributed randomly.
We run $5\times10^4$ integrations and explore what happened to the stellar binary
during the first close passage by the SMBH.
We distinguish five outcomes for the evolution of the stellar binaries:
merger, single TDE (may be accompanied by HVS), double TDE and uneventful fly-bys.
We identify merger with $a_1(1-e_1)<R_1+R_2$.
When the distance between $m_2$ and $m_\alpha$ ($\alpha=1,2$) is smaller than $r_{\alpha t}$,
we refer to this as TDE.
Additionally, if TDE occurs on either of the components of the stellar binary,
the other may be ejected and run out of the gravitational potential well of the SMBH.
When the velocity of this ejected star is higher than $500$km/s, we record a HVS here.
For the fifth category, we label the stars neither disrupted nor merger after a close approach
as uneventful fly-bys.

\subsection{Statistical Results}

\begin{figure}
\centering
\begin{tabular}{c}
\includegraphics[width=8.5cm]{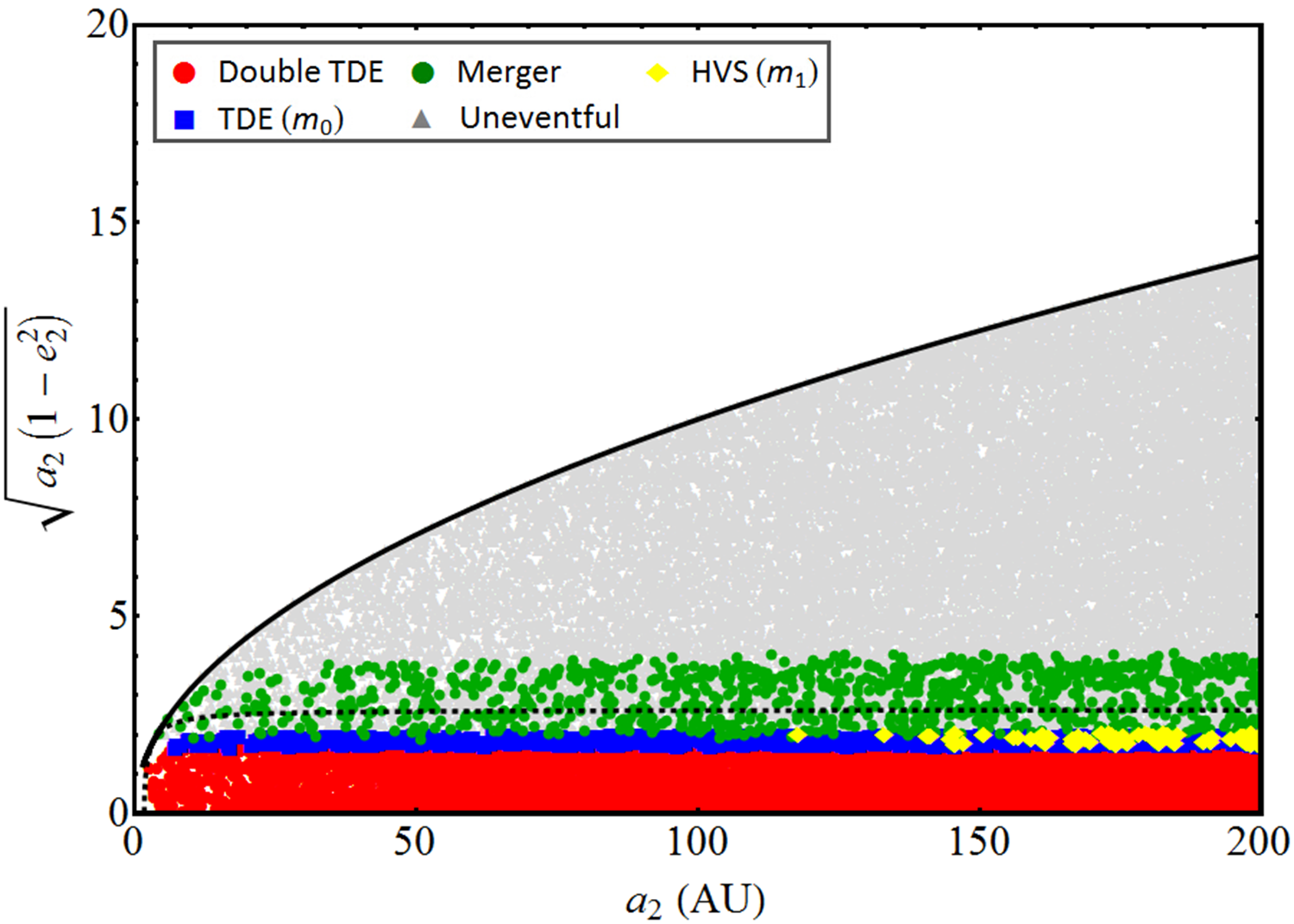}\\
\\
\includegraphics[width=8.5cm]{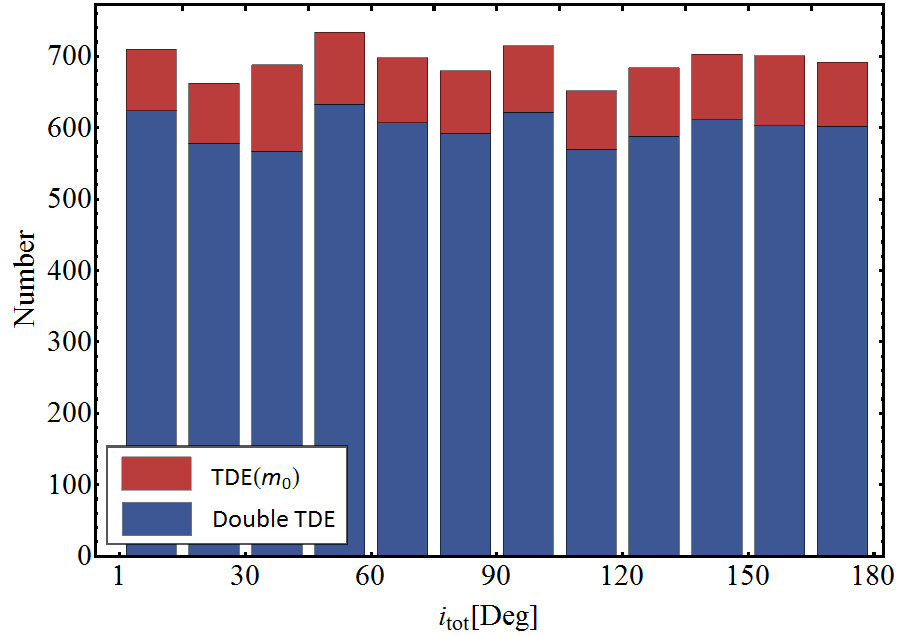}
\end{tabular}
\caption{Upper panel:
different endings of the stellar binaries with one orbit passing close to an SMBH
are shown in the stellar angular momentum-energy phase space.
Lower panel: the numbers of the single and double TDEs distributed in the initial inclinations.
}\label{fig:5}
\end{figure}

As shown in the upper panel in Fig. \ref{fig:5}, we present the destinies of stars from numerical experiments in
$\sqrt{a_2(1-e_2^2)}-a_2$ phase space, which is associated with the orbital angular momentum $L_{{\rm Bin},B}$ and the orbital energy $E_{{\rm Bin},B}$
of Binary B, respectively
\be
\begin{split}
&E_{{\rm Bin},B}\sim -\frac{Gm_2(m_0+m_1)}{2a_2},\\
&L_{{\rm Bin},B}\sim\frac{m_2(m_0+m_1)}{m_0+m_1+m_2}\sqrt{G(m_0+m_1+m_2)a_2(1-e_2^2)}.
\end{split}
\ee

Among our $5\times10^4$ simulations, 80.7 per cent are survival, 2.2 per cent are 
mergers, 2.7 per cent
are single TDEs on the more massive star ($m_0$) and
14.4 per cent are double TDEs.

In the upper panel of Fig. \ref{fig:5},
the black curve provides a boundary of the eccentricity of Binary B,
below which the phase space is indicated to have large $e_2$.
The black dashed curve given by $r_{\rm bt}=r_{\rm per}\equiv a_2(1-e_2)$ 
shows that the stellar binary approaches the SMBH with the
pericentre distance as large as the tidal separated radius.
Here, the destinies of the binary are colour coded (as labelled).
During the first close encounter, the orbital energy changes, giving chances for the binary to be break-up.
We find that the fates of the stars depend sensitively on the angular momentum of the orbit rather than the energy.

For any values of $a_2$, the angular momentum decreases when $e_2$ becomes larger.
We find that the stellar binary has more opportunities to be survival with lower $e_2$.
Meanwhile, compared to the uneventful fly-bys, the numbers of the mergers are rare but distributed
below a critical value of the angular momentum ($e_2\gtrsim0.6$).
Most of them locate outside the `Tidal Loss Cone' 
($r_{\rm per}<r_{\rm bt}$).
Although a close passage by the SMBH has provided numerous mergers,
the following simulations will reveal that the Kozai mechanism produces more after many orbits around the SMBH.

When the system enters the so-called `Tidal Loss Cone',
it is possible for either (or both) of the two stars to be tidal disrupted.
Due to the fact that the tidal radius of $m_0$ is larger than $m_1$'s,
the single TDEs recorded in these simulations are all from $m_0$.
In particular, when $e_2\gtrsim0.98$,
$13.8\%$ of the total single TDEs are accompanied by the production of HVSs of $m_1$.

However, when $\sqrt{a_2(1-e^2_2)}$ is sufficiently small and the periapsis separation $r_{\rm per}\lesssim 0.3 r_{\rm bt}$,
the stellar binary approaches the SMBH on a nearly radial orbit.
In this case, both of the stars loss the angular momentum, 
falling into individual tidal radius (equation \ref{eq:RAT}),
ending up with the double TDE \citep[e.g.,][]{DoubleTDE}

The lower panel shows the numbers of two types of TDEs of the stellar binaries as a function of $\I$,
which corresponds to the initial inclinations between two orbital planes.
We find that, single or double TDE does not have significant dependence on the orbital orientation.

\section{Numerical Exploration on the four-body system}

\begin{figure*}
\centering
\begin{tabular}{cc}
\includegraphics[width=8cm]{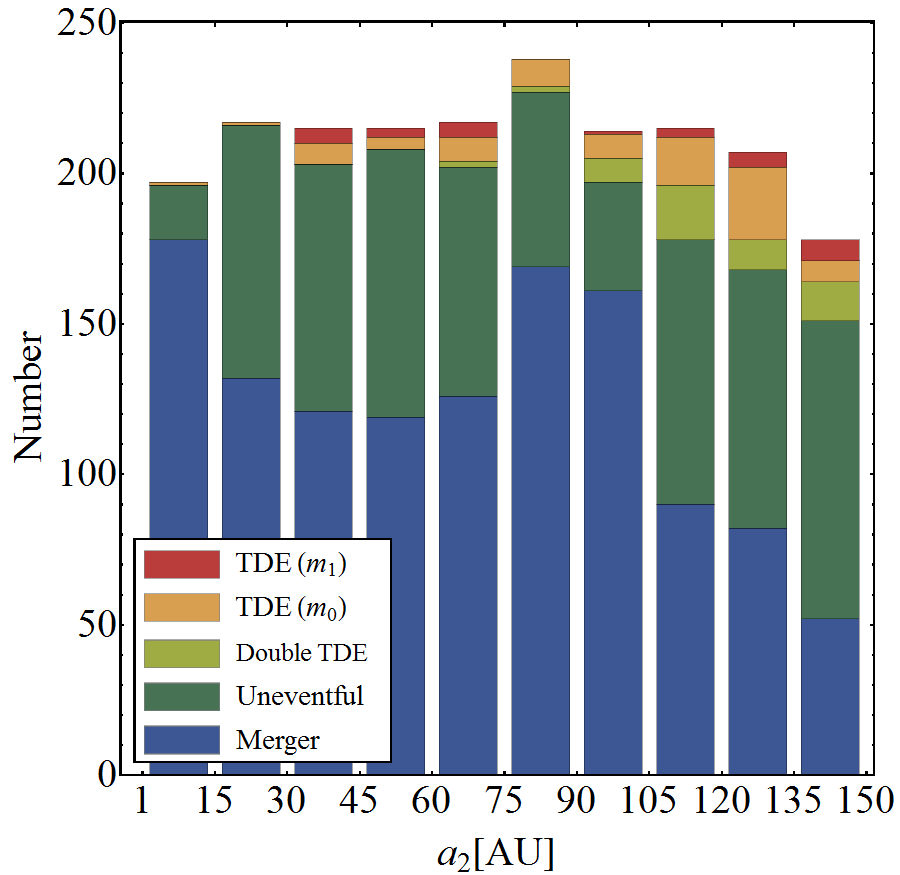}&
\includegraphics[width=7.8cm]{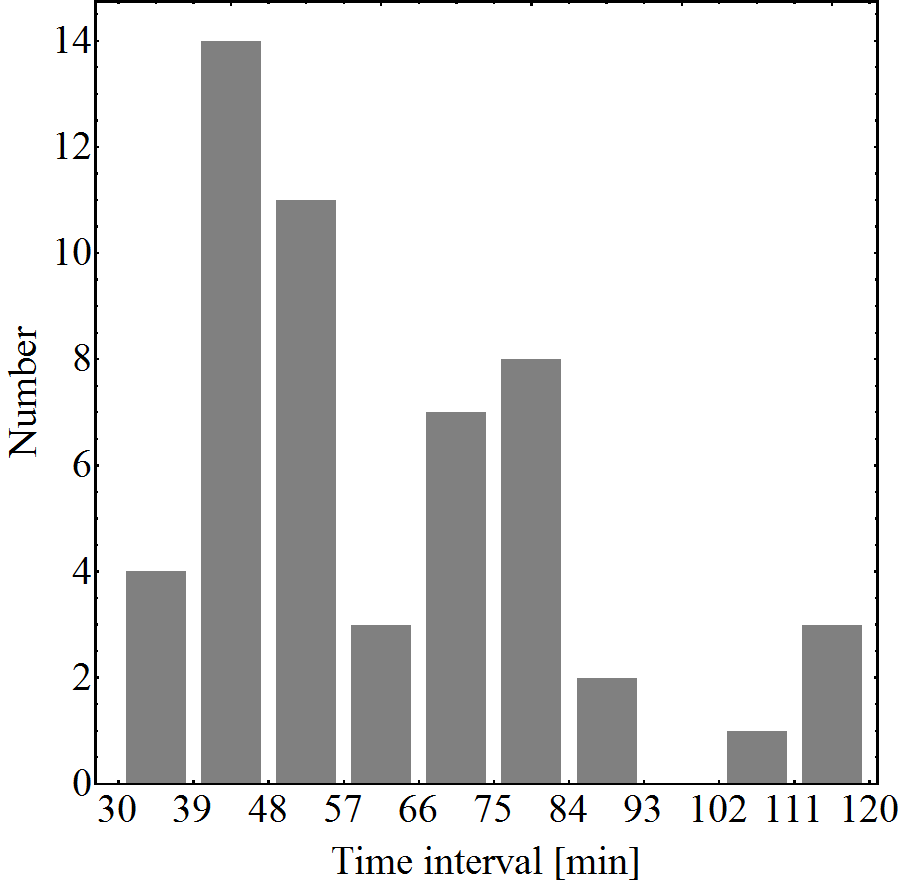} \\
\includegraphics[width=8cm]{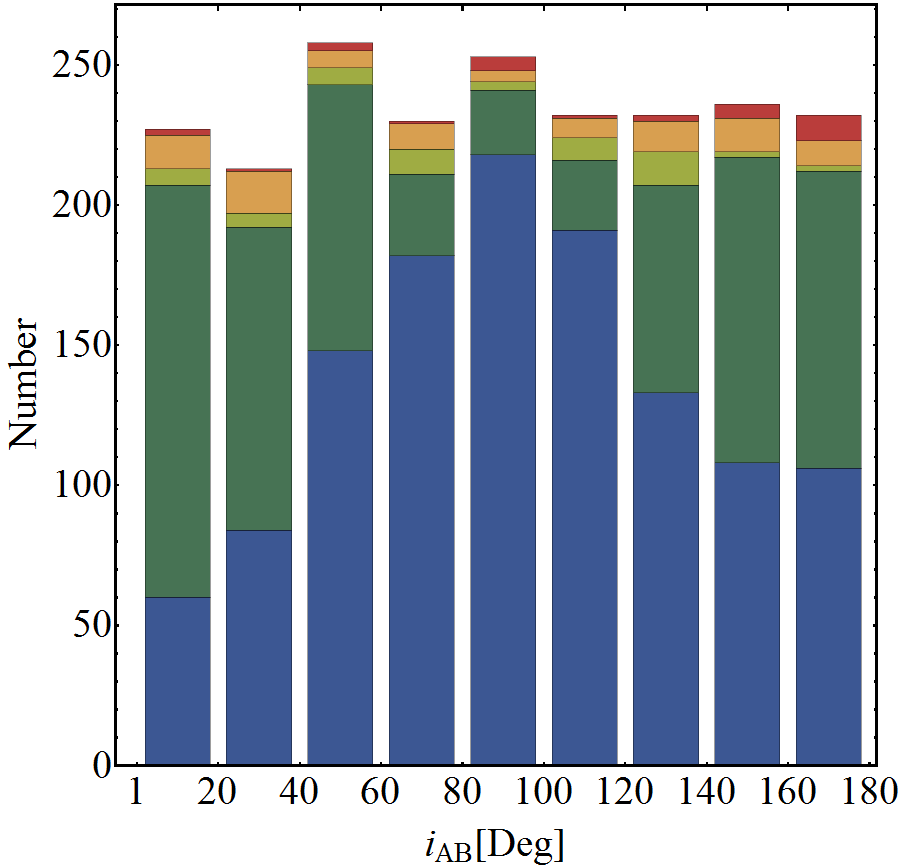}&
\includegraphics[width=8cm]{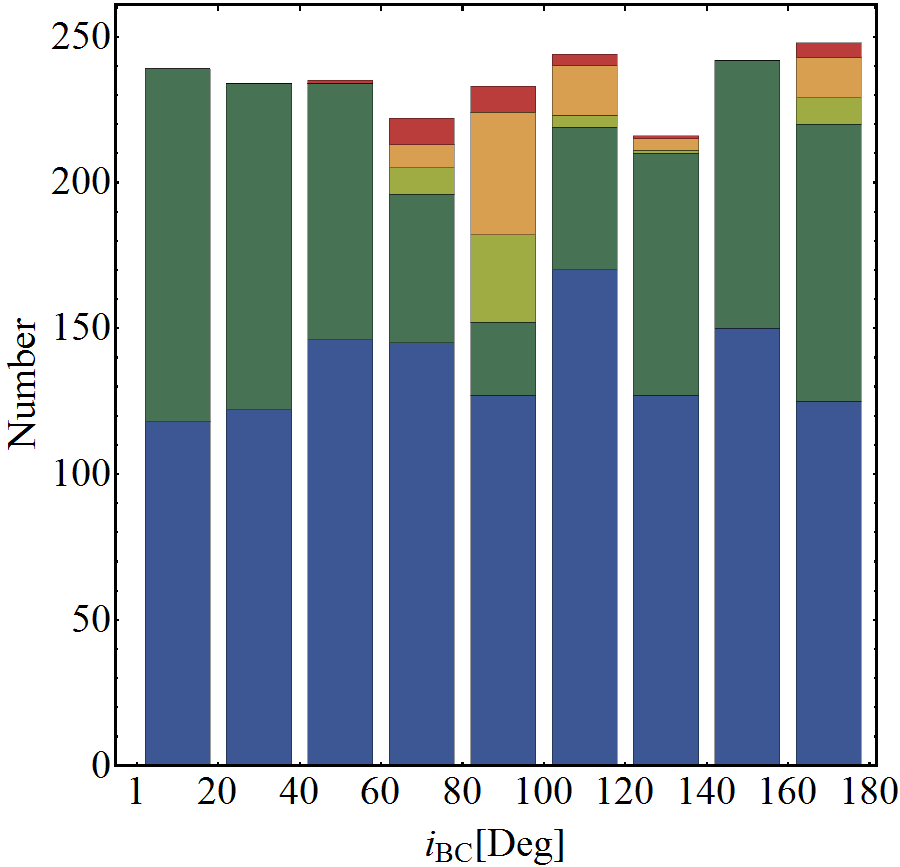} \\
\end{tabular}
\caption{The final distributions of different endings of the stellar binaries for our illustrative example:
$m_0=2 \msun$, $m_1=1 \msun$, $m_2=m_3=10^6 \msun$, $a_1=0.05\au$, $a_3=400\au$ and $e_{1,0}=e_{2,0}=e_{3,0}=0.001$,
with the initial uniform distributions in $a_2$, $i_{AB}$ and $i_{BC}$.
The top right panel shows the distributions of the time interval (between
two single TDEs) for the total double TDEs.
}\label{fig:6}
\end{figure*}

\begin{figure*}
\centering
\begin{tabular}{ccc}
\includegraphics[width=5.5cm]{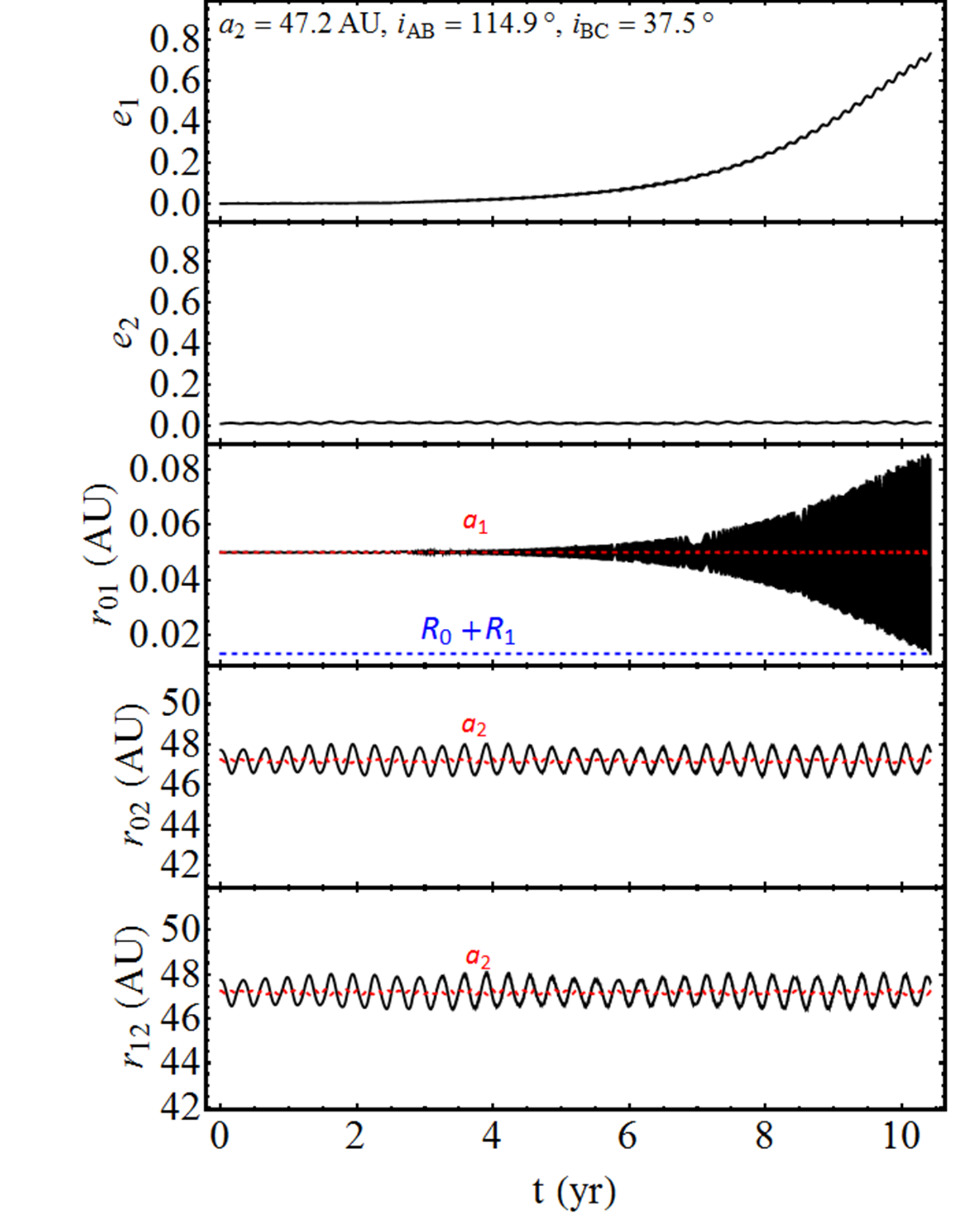}
\includegraphics[width=5.5cm]{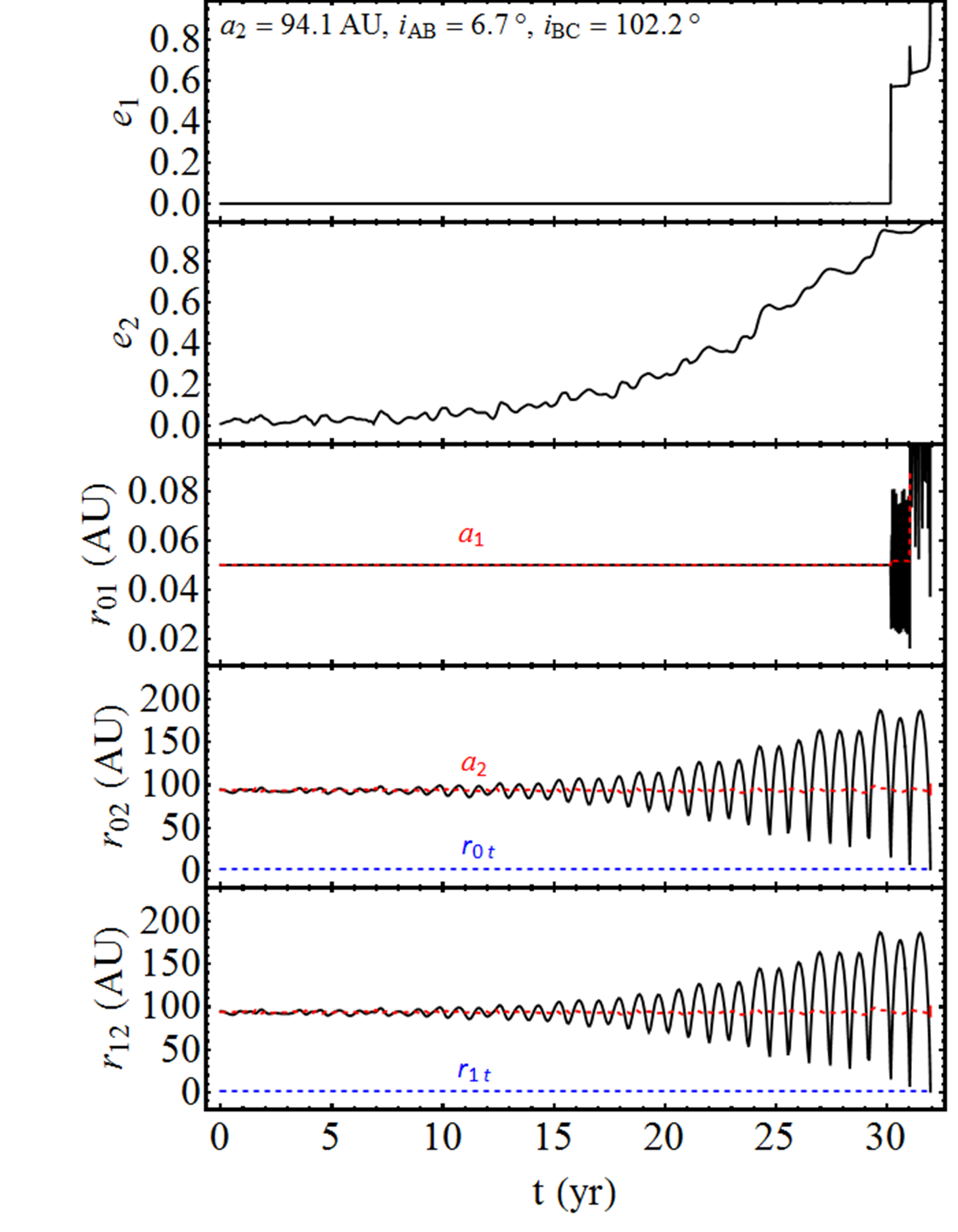}
\includegraphics[width=5.5cm]{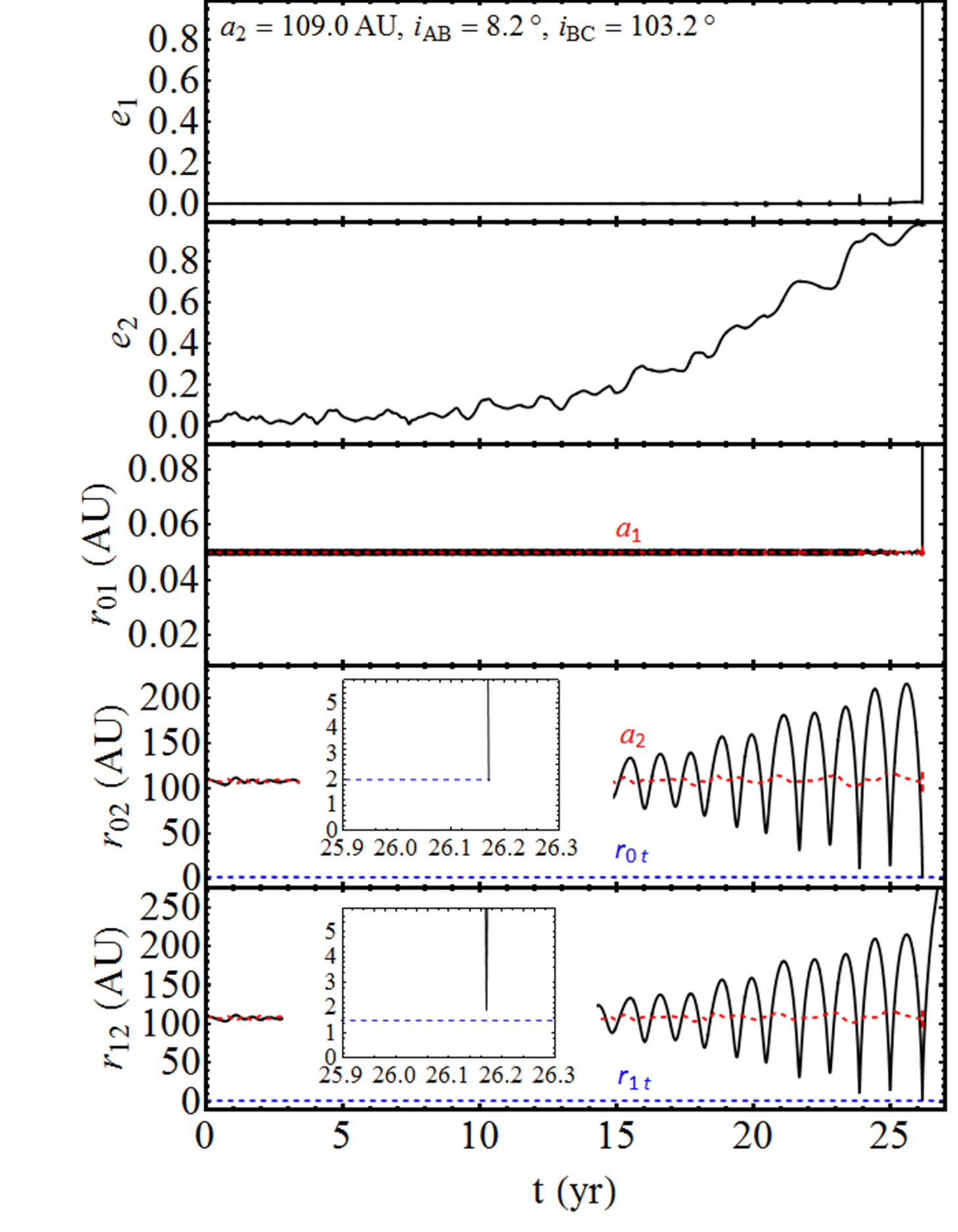}
\end{tabular}
\caption{The time evolution of the orbital elements for several cases chosen in Fig. \ref{fig:6}.
From the top to bottom, the orbital elements are the eccentricities of Binary A and B ($e_1$ and $e_2$),
and the separations between $m_0$, $m_1$ and $m_2$ ($r_{01}$, $r_{02}$ and $r_{12}$).
Here, $R_{\alpha}$ and $r_{\alpha t}$ ($\alpha=0,1$) are the radius and tidal disrupted radius of each star, respectively.
From the left to right, the outcomes of the stellar binary are the ``typical" merger, double TDE and single TDE on $m_0$.
}\label{fig:7}
\end{figure*}

\begin{figure*}
\centering
\begin{tabular}{ccc}
\includegraphics[width=5.5cm]{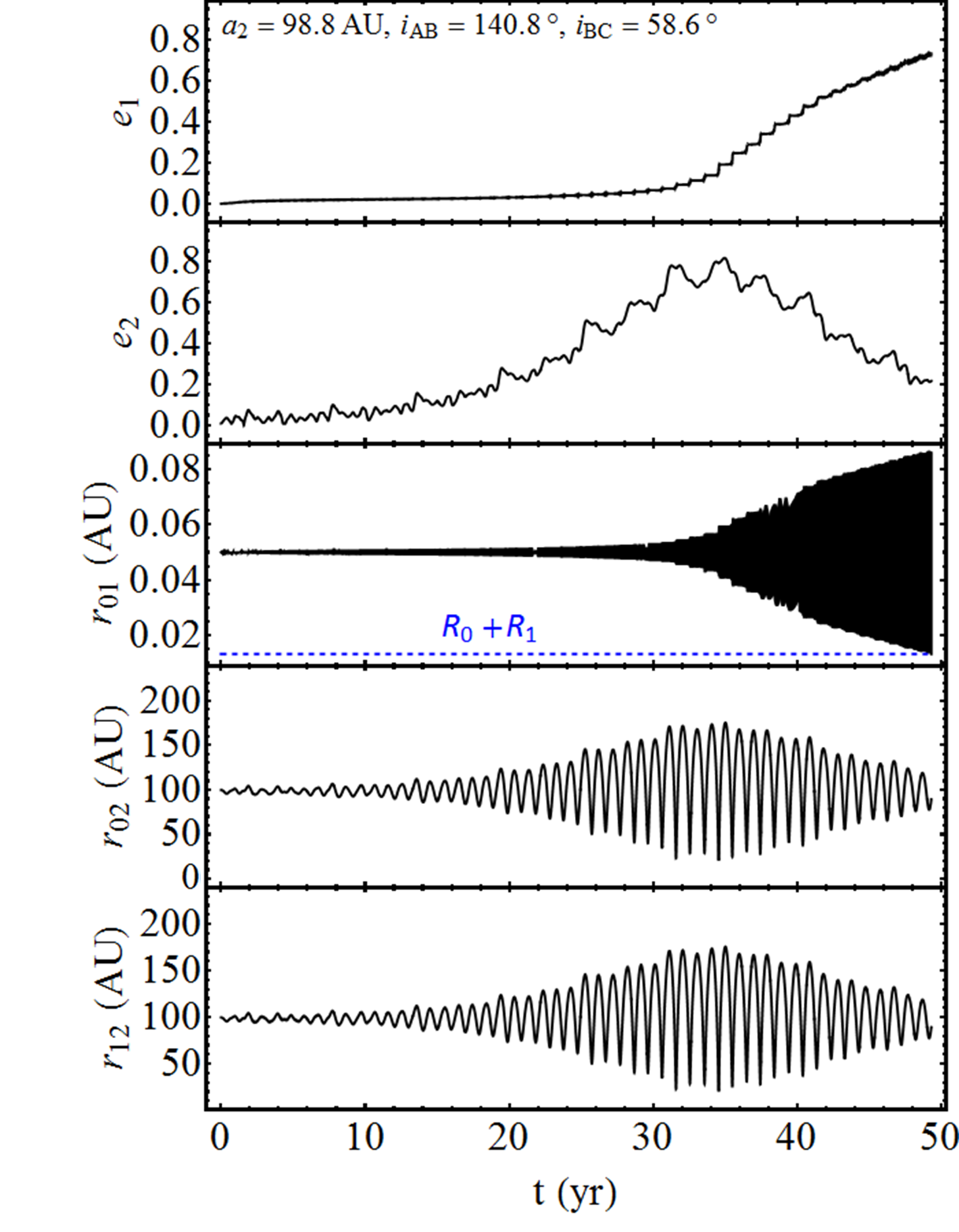}
\includegraphics[width=5.5cm]{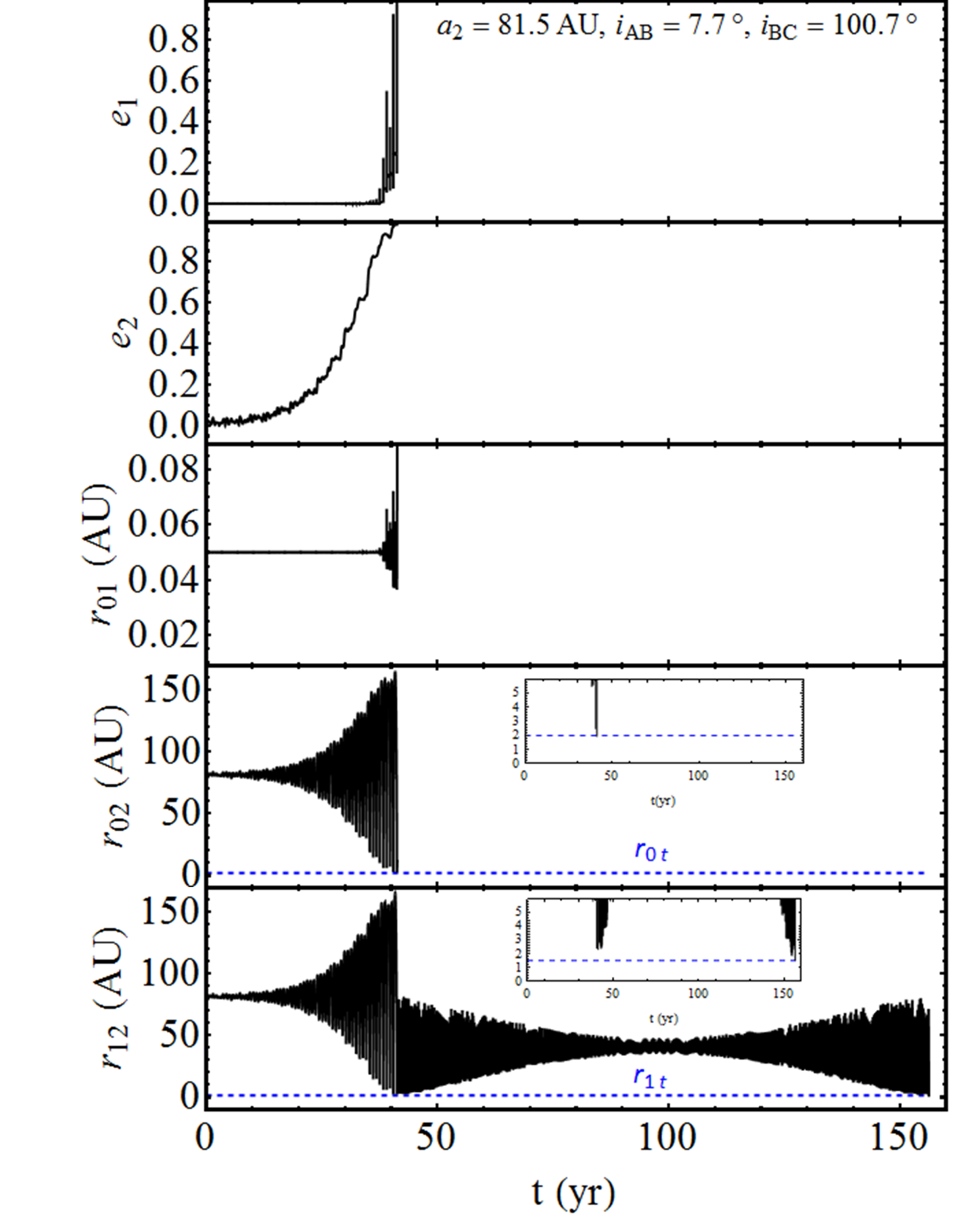}
\includegraphics[width=5.5cm]{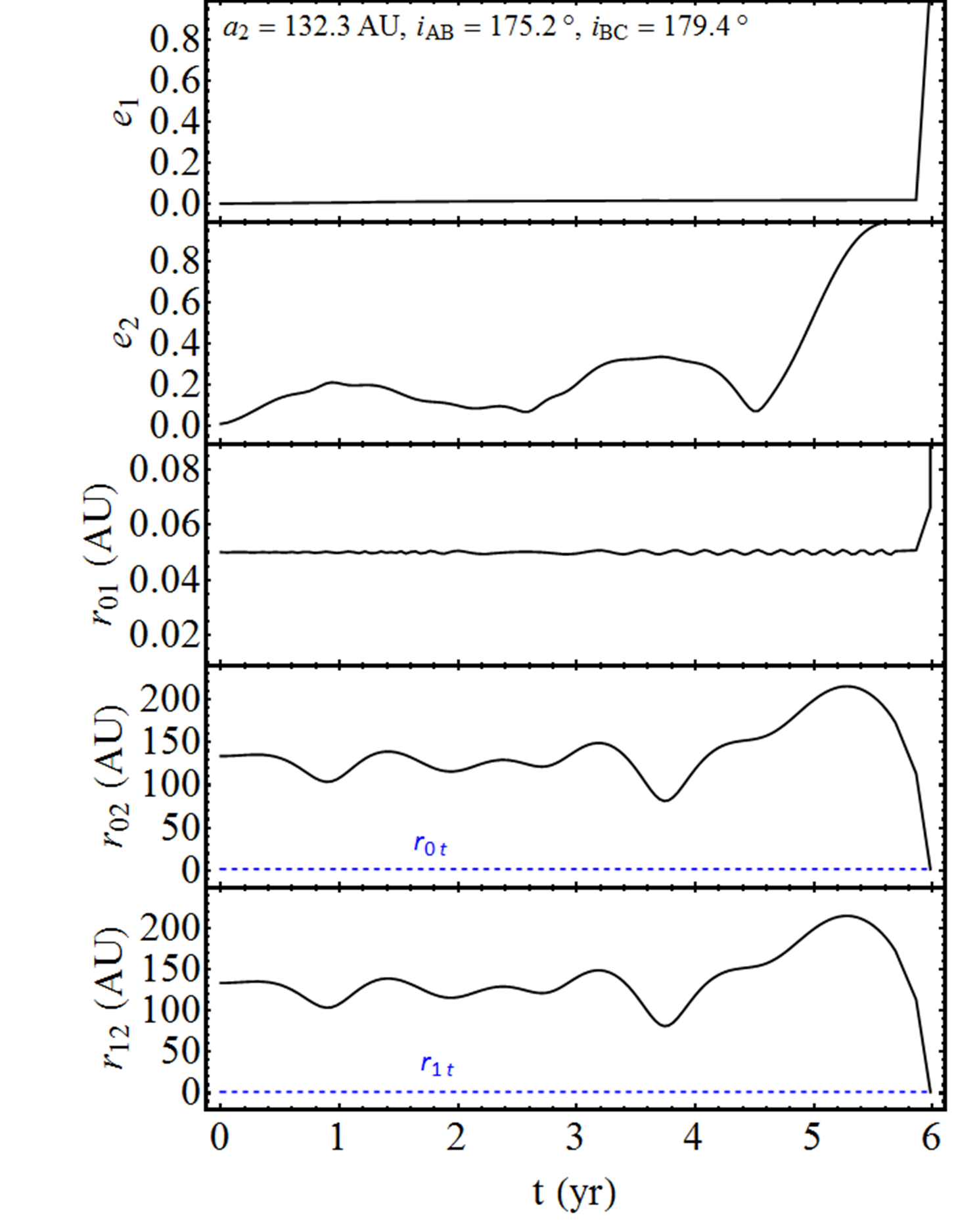}
\end{tabular}
\caption{Similar to Fig. \ref{fig:7}, but some `special' cases with the merger, single TDE and double TDE (from left to right).
}\label{fig:8}
\end{figure*}

Figure \ref{fig:5} indicates that the destinies of the stellar binaries are determined by the initial parameters (i.e., $L_{{\rm Bin},B}, E_{{\rm Bin},B}$).
Exactly, the existence of a distant SMBH may induce the LK oscillations 
in the outer triple
so that the stellar binary can approach to the first SMBH in a variety of possible orbits.
Due to the close encounter with the SMBH, the motion of the system is failed to be described by the orbital averaged analysis.
In this section, we try to explore the evolution of the stellar binary in the four-body system by carrying
out $N$-body integrations,
and discuss the process that can repopulate the binaries.

\subsection{Numerical setup}

In our illustrative example:
Binary A is consisted of two stars as $m_0=2 \msun$ and $m_1=1 \msun$ with a tight semimajor axis $a_1=0.05\au$.
Since the distrubuton of stars
surrounding SMBH binary remains unclear to date, and the current numerical
results have model dependence inevitability.
For simplicity, we focus on the binary close to the SMBH in the allowed parameter space
(shown in Fig. \ref{fig:2} and \ref{fig:3}).
The distance between the centre of mass of the stellar binary and the first SMBH is assumed to be distributed uniformly
at a short range of $a_2\sim(1\au,150\au)$.
The secondary SMBH has mass of $m_3=10^6 \msun$ and the separation of SMBH is chosen to be $a_3=400\au$.
The inner and outer triples are setup by uniform mutual inclinations ($i_{AB}$ and  $i_{BC}$) between $0^\circ$ and $180^\circ$,
and the eccentricities are set to be $e_{1,0}=e_{2,0}=e_{3,0}=0$ initially.

In particular, it is required that each component of Binary A is outside its own tidal disrupted radius $r_{\alpha t}$
at the beginning.
We run 2000 Monte-Carlo $N$-body integrations for the time-scale $\sim10(T_\kb+T_\kc)$,
without involving the short range forces and the energy dissipation.
Three outcomes are distinguished for the evolution of the stellar binaries:
merger, TDE and uneventful fly-bys.
Using the criterions listed in Section 3.1,
we also identify the double TDE if the time interval between two TDEs is shorter than $\sim 0.01/n_2$ especially
(where the two stars are expected to be tidal disrupted sequentially).
Otherwise, we refer to the single TDE, depending on either of $m_0$ and $m_1$ is tidal disrupted first.
Since we define that each set of initial parameters has only one corresponding result,
when TDE occurs on $m_0$ or $m_1$, the motion of the other is not recorded.
Actually, the surviving companion may experience the partial tidal disruption at the first encounter by the SMBH
and will be fully disrupted at the following encounters if the star is still bound to the SMBH.
Additionally, we do not include the fate of the massive star produced by the merger,
and if either of the stars is captured by the secondary SMBH we do not make a record as well.

\subsection{Statistical Results}

Figure \ref{fig:6} shows the final distributions of our $N$-body simulations with $a_3=400\au$.
Most of the events occur within $\sim(T_\kb+T_\kc)$ and the different endings of the stellar binaries
are marked by the `rainbow' colours.
Out of all, 2000 stellar binaries between $a_2=1\au$ and $150\au$, $58.2\%$ are mergers,
$7.9\%$ are TDEs (totally) and $33.9\%$ are survival.
Note that half of the events of the single tidal disruption are accompanied by the production of HVSs (not labeled).
Since the definition of the double TDE is not clear, we plot the distribution of the time interval between two single TDEs
as well, which are all satisfied with $< 0.01/n_2$.
It seems that most of the secondary TDE occurs in tens of minutes later after the first TDE.

On the basis of the above analysis, the stellar binaries with close distance from SMBH are
more likely to merge ($T_\kb<T_\kc$). If the stars surrounding SMBH are truncated at a larger semimajor axis ($a_2$),
$T_\kb$ will approach $T_\kc$ (equations \ref{eq:KozaiB}, \ref{eq:KozaiC}),
the stars have more opportunities to be tidal disrupted and the merger rate will decrease
(see the upper left panel in Fig. \ref{fig:6}).
Although the initial inclinations have an approximate isotropic distribution,
the Kozai effect plays a major role with high inclinations,
the eccentricities of the stellar binaries can be driven to be large.
So, the distribution of mergers in the final inclinations
has a strong corresponding peak
around $90^\circ$. The fraction of mergers with the inclination $i_{AB}\sim(40^\circ, 140^\circ)$ is $\sim70\%$
of the total mergers,
while the TDEs still have a thermal distribution (see the lower left panel in Fig. \ref{fig:6}).
Similarly, the outer triple with high inclinations will also undergo the LK oscillations,
making the angular momentum $L_{{\rm Bin},B}$ sufficient small.
The stellar binaries get close to the SMBH and TDE occurs,
which with $i_{BC}\sim(40^\circ, 140^\circ)$, accounting for $\sim73$ per cent of all possible inclinations
(see the lower right panel in Fig. \ref{fig:6}).
However, it is noted that the number distribution in the semimajor axis $a_2$ shows a secondary peak at the larger $a_2$,
and there are still some merger events and TDEs as shown out of 
the LK dominated region ($0^\circ - 40^\circ$ or $140^\circ - 180^\circ$).
We will explore these `unexpected' distributions in more detail further.

Figure \ref{fig:7} shows the examples of the evolution of different final endings from our 
$N$-body simulations,
which presents the corresponding orbital elements as the eccentricities of Binaries A and B ($e_1$ and $e_2$), 
the separations between $m_0$, $m_1$ and $m_2$ ($r_{01}$, $r_{02}$ and $r_{12}$) in different columns.
As expected in the parameter space (Fig. \ref{fig:3}), the representative stellar mergers occur for $T_\kb<T_\kc$ and TDEs
are inserted in the region of $T_\kb>T_\kc$,
where the LK oscillations dominate the inner or outer triples, respectively.
Noted that, in the middle panel,
the angular momentum of the stellar binary get a ``kick" during the sufficient close passage by the SMBH,
exchanging the orbital energy. The eccentricity rapidly jumps to a high value.
In the subsequent evolution, the binary components are still bound to each other until the second periapsis passage.
After that, the stars will be separated, leading TDEs or experience clean collision.
This has also been pointed out by \citet{Antonini 09}.

However, we also find that the mergers are able to occur even if the Kozai effect operates in the outer triple initially.
As shown in the left-hand panel of Fig. \ref{fig:8},
for the relatively distant stars, the pericentre distance of them can be closer by the outer Kozai oscillations.
Therefore, the inner Kozai effect comes into play with high $e_2$ and the eccentricity of the stellar binary is excited
in a smooth-like way.
This is also associated with the secondary peak of the distribution in $a_2$ in the top left panel of Fig. \ref{fig:6}.
In addition, we find such a type of systems, where one component of the binary undergoes tidal disruption and
the surviving companion is still bound to the SMBH in a highly eccentric orbit after a close encounter.
But, because of the existence of the outer SMBH,
the star may experience the Kozai oscillation induced by this newly produced triple system (SMBH-star-SMBH),
exchanging energy at periastron, and TDE occurs eventually \citep[]{Li TDE}
(as shown in the middle panel of Fig. \ref{fig:8}).
Since the time interval between two TDEs is so long that beyond $\sim 0.01/n_2$,
we identify this event as the single TDE rather than the double TDE.
Finally, to check the ``unexpected" distribution we choose several cases to
explore the time evolution of the orbital elements (as shown in the right-hand
 panel of Fig. \ref{fig:8}).
In such an example,
the system is appeared to be very unstable. As a result, the quantities are varying irregularly.
There are still some other non-hierarchical systems that contribute the distributions in the inclinations near $0^\circ$ or $180^\circ$.

\subsection{Efficiency of Double Kozai Oscillations}

\begin{figure}
\centering
\begin{tabular}{ccc}
\includegraphics[width=8cm]{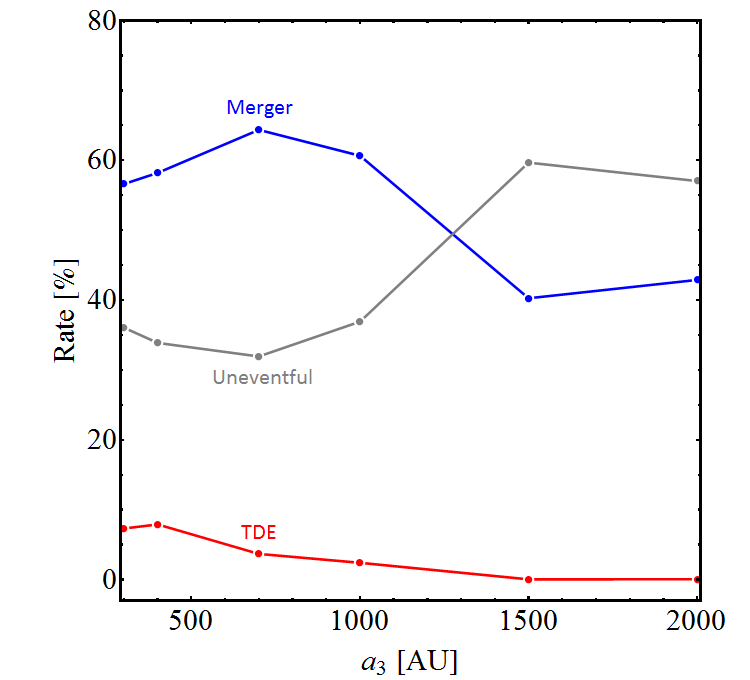}
\end{tabular}
\caption{The variations of the rates of the stellar mergers, TDEs and uneventful events
as a function of different semimajor axis of the SMBH binary.
}\label{fig:9}
\end{figure}

The Kozai effect in the outer triple is no longer to be important if the secondary SMBH is very far away.
Therefore, the evolution of the stellar binary is determined by the first SMBH completely.
Inversely, if $m_3$ becomes closer, the outcome of the main sequence binary is influenced by both of the LK oscillations.
In order to detect the significance of the external SMBH,
we tune the semimajor axis $a_3$ gradually in this section and find out what happens to the rates.
Again, we use the same initial parameters and distributions as the previous sections.
For each $a_3$, we carry out 2000 integrations for the time-scale $\sim10(T_\kb+T_\kc)$.

Figure \ref{fig:9} depicts the rates of different events as a function of $a_3$.
In comparison to the mergers, TDEs have a relatively small rate.
However, it is a natural consequence of the LK oscillation induced by the secondary SMBH.
We also find that the enhanced rate of mergers achieves its maximum at $a_3\sim700\au$,
that is $\sim1.5$ times higher than the one without the second SMBH,
whereas the rate of TDEs has a peak around $a_3\sim400\au$.
As we discussed in the comparison between the Kozai time-scales from the inner and outer three-body (Fig. \ref{fig:3}),
for a given range of $a_2$ $(1\au-150\au)$, TDEs are predicted to be the most of the products when $a_3\sim550\au$.
On the other hand,
the maximum probability of occurrence of a merger event is supposed to occur with $a_3\sim1000\au$.
Note that all the rates are significantly decreased when $a_3\sim2000\au$,
and the affect of the outer SMBH is negligible typically.
Of course, because the gap in $a_3$ we have chosen is large,
the extreme values of the rates are not traced accurately.
However, the overall conclusion is held qualitatively.

\section{Summary and Discussion}

SMBH binary is expected to be the by-product of the merging galaxies.
In this work, we consider the binary main sequence stars  orbit around the SMBH,
perturbed by a distant SMBH.
Due to the existence of the secondary SMBH,
the evolution of the stellar binary varies a lot through LK oscillations or the
chaotic interactions.
The rates of different final endings of the main sequence stars are affirmed to be enhanced significantly,
compared to the case without the outer SMBH.
Due to the enhanced merger rate of the stellar binaries,
the population of the massive stars (OB star) will be modified \citep[]{Perets and Fabrycky}.
Similar process could also take place for the compact objects, increasing the formation of Type Ia supernova,
gamma ray burst and gravitational wave sources.
Contrarily, the increased rates of the events can predict the existence of the external SMBH,
that is difficult to be detected as mentioned in \citet{liufukun} and \citet{Li TDE}.

Comparing the time-scales of various processes related to the galactic center,
we point out a non-neglected regime in the parameter space where the Kozai effect induced by the second SMBH is strong enough.
Due to the possible LK oscillations in the outer triple (SMBH-stellar binary-SMBH),
the stellar binary has more chances to experience a close passage by the SMBH.
If the pericentre is close enough,
the dynamics of the stellar binary is supposed to be diverse.
We choose an illustrative example to study the evolution of this four-body system.
By performing $N$-body integrations,
our simulations demonstrate that the rate of mergers between the stars is $64.4\%$ for a distant SMBH binary ($a_3=700\au$),
that is $\sim1.5$ times higher than the one without the external SMBH,
whereas the rate of TDE is as many as 7.9 per cent when $a_3=400\au$
(this rate is expected to be zero without the perturber).
Therefore, the LK effect produced by the secondary SMBH is the major driver for the occurrence of the TDEs.
Additionally, the final ending of the stellar binary is determined by the initial parameters of its external orbit
around the first SMBH
that can be produced by the LK oscillations in the outer triple.
Relative to the orbital energy, we find that the outcomes of the stars
depend sensitively on the angular momentum of the orbits.
In general, the orbit of the stellar binary with low angular momentum has more opportunities to
be tidal disrupted and the opposite is more likely to survive or merge.

There are a few uncertainties in our estimation on the rates before making a firm statement.
Foremost, we perform our calculations with uniform distribution of the stellar binaries
(both in the inclinations between different orbital planes and the distances to the SMBH)
and assume a fixed semimajor axis and masses of the stellar binary (that may change the Kozai time-scale) in Section 4.
However, in the `realistic' astronomical systems, such as inspiral or merger of the galaxies \citep[e.g.,][]{Ibata}
and the massive BH binary \citep[e.g.,][]{Makino},
the stellar binaries are distributed unclearly.
Since the dynamical friction plays a fundamental role between the SMBH binary and the stellar binaries,
the frictional force may affect the stellar distribution surrounding the SMBH,
reducing the density described by equation (\ref{eq:density}) \citep[e.g.,][]{Antonini 11}.
Also, these `background' stars try to tighten the SMBH binary via the sling-shot mechanism
\citep[e.g.,][]{Roos 1981,Roos 1988}.
Consequently, a more sophisticated stellar distribution is required through the numerical experiments that
couples the evolving SMBH binary with stellar scattering.
After that, the numbers of events would be produced by special stellar distribution with varied distance.
Besides, according to Fig. \ref{fig:6}, the possibilities of different fates can still be estimated roughly.

In addition to the distribution of the stellar binaries,
the evolution of the system can be modified by the parameters of the secondary SMBH.
Such as, the mass of SMBH is in a range for $10^6\msun -  10^9\msun$ at least.
Based on equation(\ref{eq:KozaiC}),
the massive perturber may reduce the Kozai time-scale,
increasing the efficiency of it.
Especially, if the SMBH binary has an elliptical orbit,
the octupole effect becomes important, driving the eccentricity to be extremely large.
Therefore, the stellar binaries will have more opportunities to be sufficiently close to the SMBH
than the one at the quadrupole approximation.
We will focus on this study in the future work.

\section*{Acknowledgements}

The authors thank the anonymous referee for suggestions of the manuscript.
BL thanks Dong Lai, Su-Bo Dong and Jun-Tai Shen for some helpful discussions.
This work is supported in part by grants from 
the National Natural Science Foundation
(No. U1431228, No. 11133005, No. 11233003,
No. 11421303).

\label{lastpage}

\end{document}